\documentclass{ecai}  

\usepackage{graphicx}
\usepackage{latexsym}
\usepackage{comment}
\usepackage{colortbl}
\usepackage{balance}
\usepackage{soul}
\usepackage{booktabs}
\usepackage{graphicx}
\usepackage{amstext}
\usepackage{multirow}
\usepackage{multirow, makecell}
\usepackage{hyperref}

\usepackage[boldmath]{numprint}
\usepackage{algorithmicx}
\usepackage{pdflscape}
\usepackage{amssymb}
\usepackage{rotating}
\usepackage[table, x11names]{xcolor}

\usepackage{bbding}
\usepackage{amsmath}

\usepackage{algpseudocode,algorithm}
\algnewcommand\algorithmicforeach{\textbf{for each}}
\algdef{S}[FOR]{ForEach}[1]{\algorithmicforeach\ #1\ \algorithmicdo}

\npdecimalsign{.}
\nprounddigits{2}

\begin{document}
	
	\begin{frontmatter}
		
		\title{Multi-Value Alignment in Normative Multi-Agent System: An Evolutionary Optimisation Approach}
		
		\author[A]{\fnms{Maha}~\snm{Riad}\orcid{0000-0003-1128-2579}\thanks{Corresponding Author. Email:maha.riad@ucdconnect.ie.}}
		\author[B]{\fnms{Vinicius}~\snm{de Carvalho}\orcid{0000-0002-4623-7244}}
		\author[A]{\fnms{Fatemeh}~\snm{Golpayegani}\orcid{0000-0002-3712-6550}} 

		\address[A]{School of Computer Science, University College Dublin (UCD)}
		\address[B]{Escola Polit\'ecnica (EP), University of Sao Paulo (USP)}
		
		\begin{abstract}
			Value-alignment in normative multi-agent systems is used to promote a certain value and to ensure the consistent behaviour of agents in autonomous intelligent systems with human values.
			However, the current literature is limited to the incorporation of effective norms for single-value alignment with no consideration of agents' heterogeneity and the requirement of simultaneous promotion and alignment of multiple values. 
			This research proposes a multi-value promotion model that uses multi-objective evolutionary algorithms and decentralised reasoning to produce the optimum parametric set of norms that is aligned with multiple simultaneous values of heterogeneous agents and the system.
			To understand various aspects of this complex problem, several evolutionary algorithms were used to find a set of optimised norm parameters considering two toy tax scenarios with two and five values are considered. The results are analysed from different perspectives to show the impact of a selected evolutionary algorithm on the solution, and the importance of understanding the relation between values when prioritising them. 
		\end{abstract}
		
	\end{frontmatter}
	
	\section{Introduction}\label{sec1}
	Normative multi-agent systems (NorMAS) have been used effectively to coordinate the behaviour of agents in multi-agent systems (MAS) that model complex applications such as intelligent transport systems \cite{morris2019norm}.
	The norms in NorMAS are regulative norms defined by a social group to regulate behaviours~\cite{riad2022run}. For example, in a traffic system, the norm is to give priority to emergency vehicles. Also, it is the norm that passengers leave front seats in buses for senior people.
	These examples represent guidelines that might be recommended in some societies, obligated, or prohibited ~\cite{riad2022run}, so when the agents are aware of the norms of the environment they are operating in, they can synchronise their behaviour with other agents, facilitate group decision making and collaborate. However, it is essential to promote human values in MAS as well to reflect real applications.
	
	In the context of this research, the term 'value' refers to motivational values, which represent standards that serve desirable objectives \cite{sierra2019value}. In other words, a 'value' will represent a preferred state\cite{sierra2019value}, such as equality, health, fairness, etc. \cite{bench2017norms}. For example (to differentiate between a 'norm' and a 'value'), as a norm, companies give their employees maternity leaves if they have a newborn baby. However, if the values of one of the companies support equality between men and women, both can have equal maternity leaves \cite{sierra2019value}. 
	
	The concept of value-alignment was introduced in \cite{originalProblem},\cite{serramia2020qualitative} and \cite{sierra2019value}, to reflect the alignment of norms and values. Researchers used several techniques to address this challenge, they included: reasoning strategies \cite{bench2017norms}, learning methodologies \cite{noothigattu2019teaching}, utility-based approaches\cite{heidari2019agents}, and genetic algorithms \cite{originalProblem}. However, the proposed solutions neglect one or more of the following points. First, they match the norms with only one value or with the preferred sub-set of values, while in the real world, all the values 
	need to be aligned with the norms. Second, these models might not consider heterogeneous MAS, in which different groups of agents support different values, especially when these values are incompatible. For example, supporting both fairness and equality may be conflicting, as ensuring fairness does not necessarily support equality. Third, some works directly derive norms from the values of the system. However, in many systems norms and values may be incompatible and they should be considered independently. For instance, a community can have a value of supporting equality, at the same time of having a norm of giving priority to senior people in queues, or exempting them from paying taxes.
	In this paper, we address these limitations by proposing \textbf{N}orms \textbf{O}ptimisation and \textbf{V}alue \textbf{A}lignment Model (NOVA), that has three main goals:
 \vspace{-0.2cm}
	\begin{enumerate}
		\item Choose the \textit{best set of norms} in NorMAS with heterogeneous group of agents.
		\item \textit{Optimise multiple values} in NorMAS, these values can be:
   \vspace{-0.1cm}
		\begin{itemize}
			\item compatible and \textit{incompatible} values.
			\item defined by \textit{heterogeneous} groups of agents.
		\end{itemize}
   \vspace{-0.1cm}
		\item Align independent sets of norms and values in NorMAS.	
	\end{enumerate}
  \vspace{-0.2cm}
	To reach these goals,  we formalised the problem as a multi-objective optimisation problem (MOP), in which we represented the values by objectives that need optimisation, and modelled the norms as the decision variables. This allowed us to get the \textit{best set of norms} (the decision variables) when the values (the objectives) are optimised, and so, aligned the norms and the values. Moreover, solving it as a multi-objective optimisation problem, (i) allowed the system to facilitate \textit{Optimising multiple values} defined by \textit{heterogeneous} groups of agents, and (ii) allowed multiple compatible and \textit{incompatible} values (objectives) \textit{optimisation}.
	
	We proposed to solve this problem using multi-objective evolutionary algorithms (MOEAs) as they have been successfully applied to solve MOPs~\cite{surveyMetaHeuristics} in several domains including logistics, ride-sharing~\cite{deCarvalhoFatemeh2022},  environmental/economic dispatch (EED) problems~\cite{qu2018survey}, feature selection for machine learning problems \cite{coelho2021review}, and by optimising antibiotic treatments~\cite{ochoa2020multi}. 
	We applied several MOEAs (NSGA-II, MOEA/DD, SPEA2, and MOMBI2) on different evaluation scenarios to analyse the performance of each of the MOEAs.
	
	Also, as the MOEAs produce sets of non-dominant optimum solutions, to choose a final solution we extended the agents' logic with a reasoner that allowed them to vote for their preferred solution.

	Accordingly, our proposed model NOVA, is a multi-value promotion model that uses multi-objective evolutionary algorithms to produce an optimum parametric set of norms that is aligned with 
 the values of heterogeneous agents' groups. We evaluated NOVA 
 using different scenarios that measure the effect of using different combinations of values. Our contribution is three-fold:
  \vspace{-0.2cm}
	\begin{itemize}
		\item Multiple values alignment: we show the capability of choosing the optimum values for a parametric norms set while aligning it with a set of multiple optimised values.
		\item Incompatible and compatible values alignment: we model the problem as a multiple-objective problem to enable the simultaneous optimisation of all values regardless of their compatibility.
		\item Heterogeneous agents groups' values alignment: we align values from different heterogeneous groups of agents while considering shared system values.
	\end{itemize}
  \vspace{-0.2cm}
	\section{Problem Formulation} \label{section3}
	Let us consider a heterogeneous normative multi-value multi-agent system that is composed of a finite set of regular agents as $Ag=\{ag_1,ag_2,...,ag_n\}$. Each agent $ag_i$ has a set of values $V_{ag_{i}}$, a set of properties $Pr_{ag_{i}}$, a set of actions $A_{ag_{i}}$, and a set of adopted norms $N_{ag_{i}}$. There is one regulative agent $r$ that is responsible to synthesise the norms set $N$, in which $N_{ag_{i}} \subseteq N$. The norms are parametric norms, i.e. each norm $n_j$ has a set of parameters $P_{n_{j}}$ that can contain unbounded or constrained elements with discrete or continuous domains. The regulative agent $r$ has a set of values $V_{r}$ as well.
	In each step (iteration) $t$, each regular agent $ag_i$ performs actions from $A_{ag_{i}}$ and applies its set of adopted norms $N_{ag_{i}}$. The regulative agent also applies actions chosen from its set of actions $A_{r}$. Corresponding to the agents' new situations, a global state $s_t$ is captured by $r$.
	In such a system, $r$'s  main challenges are: to synthesise the optimum set of norms that ensures the alignment of its own values $V_{r}$ and each of the regular agent's values $V_{ag_{i}}$ (which is shared between a subset of agents), and to optimise the synthesised set of norms even in case of incompatible values. 
	
	\subsection{Defining the Problem as a Multi-Objective Optimisation Problem (MOP)}
	
	As the main aim is to find the best set of parametric norms when the agents' and system's values are satisfied (optimised), we consider the problem as a multi-objective optimisation problem (MOP).
	
	Multi-objective optimisation requires finding solutions which simultaneously consider two or more conflicting objectives to be minimised or maximised \cite{nagymulti}. Thus, the optimisation process aims to find a set of solutions that reflects a trade-off between the objectives. MOPs are formulated using: objective functions, constraints, decision variables and their bounds \cite{nagymulti}. 
	
	Respectively, in NOVA, we formulate the problem identified in Section \ref{section3} as a multi-objective optimisation problem. We define the agents' and system's values as the objective functions to be optimised, and the norms as the decision variables.

	\section{Norms Optimisation and Values Alignment Model (NOVA)}\label{sec:Model}
\begin{figure}[!htbp]
		\centering
		\centerline{\includegraphics[scale=0.2]{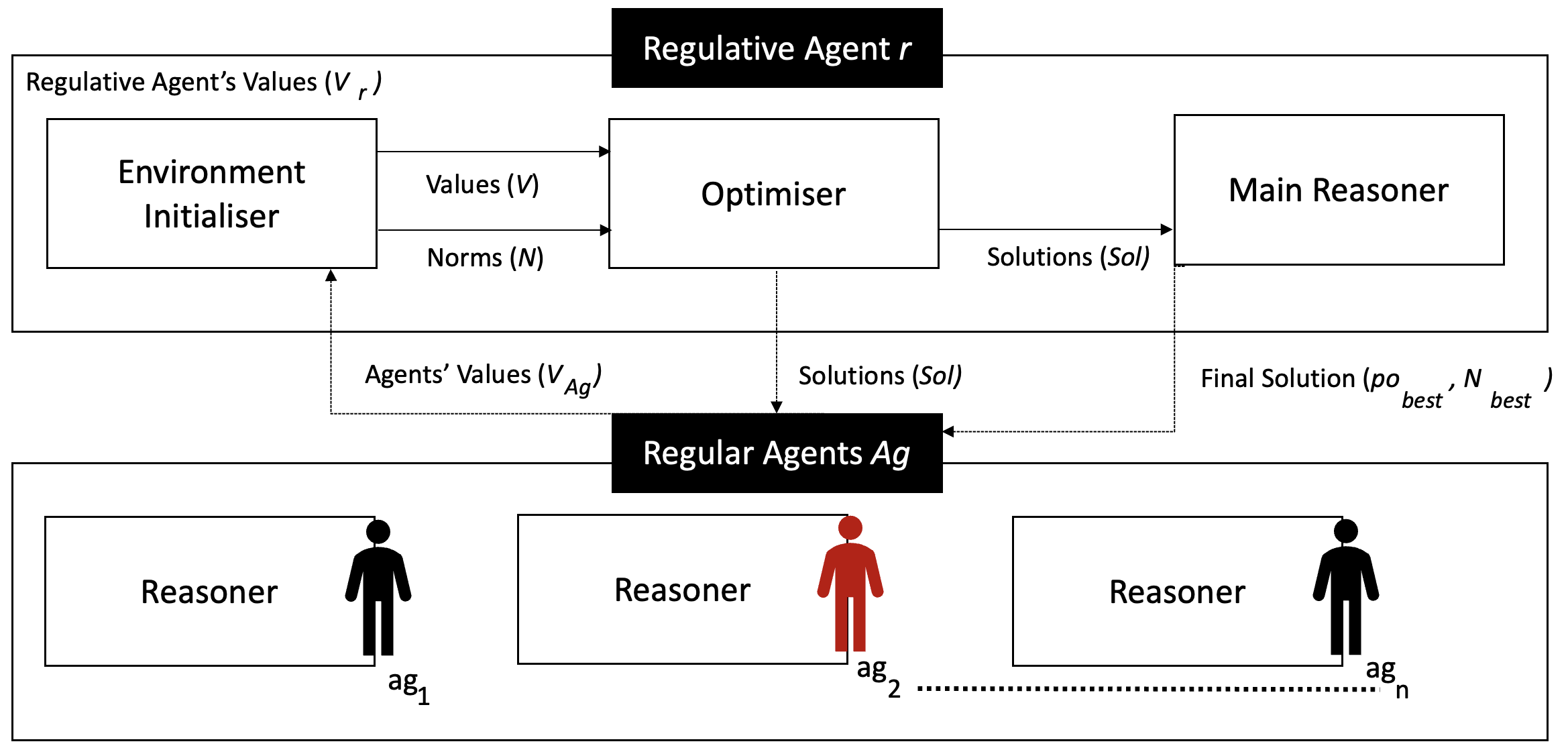}}
		\caption{NOVA Conceptual Model}
		\label{fig:GeneralConceptualModel}
	\end{figure}

	NOVA is a model for norms optimisation and values alignment, its main responsibilities are to: (i) optimise the values (objectives), (ii) choose the best set of norms, (iii) reason non-dominant solutions, and (iv) produce one final optimum solution for aligning multiple norms and objectives. 
 As seen in Figure~\ref{fig:GeneralConceptualModel}, NOVA operates using a main regulative agent $r$ and regular agents $Ag$. The regulative agent is responsible for: initialising the environment parameters and norms, collecting values from regular agents, and doing the optimisation process using the \textit{Optimiser} component. After the \textit{Optimiser} produces the set of non-dominant solutions, the \textit{Main Reasoner} in $r$ is triggered to start the reasoning process with the regular agents $Ag$.
	
	In the next subsections (\ref{subsec:optimisationProcess} and~\ref{subsec:reasoningProcess}), we illustrate  in more details the two main processes carried by NOVA.
	\subsection{Optimisation Process}\label{subsec:optimisationProcess}
	The optimisation process is performed by the \textit{Optimiser} in the regulative agent $r$ after the \textit{Environment Initialiser} (i) defines the set of parametric norms and their values bounds (ii) collects values of all agents $V_{ag}$ and integrates them with its values $V_r$ in a single set of values $V$. (iii) maps the norms as decision variables and the values as the objectives and send them to the \textit{Optimiser}. 
	
\begin{figure}[!htbp]
\centering
\centerline{\includegraphics[scale=0.285]{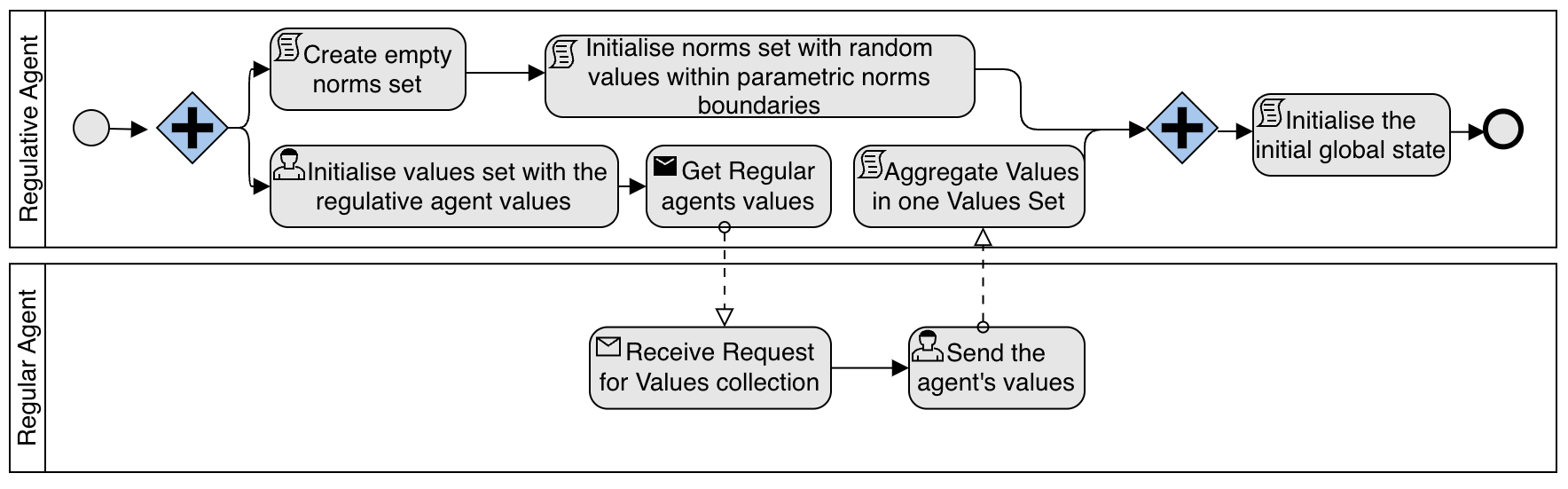}}
\caption{System Initialisation Process Instantiated by the Regulative Entity}
\label{fig:InitialisationProcessss}
\end{figure}

	Accordingly, NOVA synthesises the best set of norms (decision variables) based on the agents' values (objective functions) optimisation, using the following off-line approach:
	
	\textbf{Environment Initialisation: }initially, the parametric norms set $N$ is initialised with random values between the norms' specified boundaries by the regulative agent $r$ (as seen in Fig. ~\ref{fig:InitialisationProcessss}). Also, the primary values of the Properties $Pr_{ag_{i}}$ for each of the agents are defined. The regulative agent $r$ sends the initial set of norms $N$ to the agents. The values of the set of agents $Ag$ in the system and the regulative agent $r$ are consolidated by $r$ in one set of values $V$. These initial values are used to calculate the global state $s_0$. Then, the processed  $N$, $V_r$, $s_0$, and $MOEA$ (which is the type of the multi-objective evolutionary algorithm that will be used for the optimisation, as illustrated in the next paragraph) are used as input parameters to start the \textit{NOVA optimisation strategy} used in Algorithm 1. 

\textbf{MOEA usage in NOVA.} NOVA uses different Multi-Objective Evolutionary Algorithms (MOEAs) to solve the multi-objective optimisation problem, and produce the Pareto Front set of solutions (set of non-dominant solutions). MOEAs are heuristic techniques that provide a flexible representation of the solutions and do not impose continuity conditions on the functions to be optimised. Moreover, MOEAs are extensions of Evolutionary Algorithms (EAs) for multi-objective problems that usually apply the concepts of Pareto dominance~\cite{pdfRefOlacir}. In Pareto dominance, a certain solution $sl_a$ in the decision space of a MOP is superior to another solution $sl_b$ if and only if $f(sl_a)$ is at least as good as $f(sl_b)$ in terms of all the objectives and strictly better than $f(sl_b)$ in terms of at least one single objective. Solution $sl_a$ is also said to strictly dominate solution $sl_b$~\cite{pdfRefOlacir}.
	In Nova \textit{Optimiser}, we use four MOEA algorithms: \textbf{NSGA-II}~\cite{nsgaII}, \textbf{SPEA2}~\cite{spea2} and \textbf{MOMBI2}\cite{mombiII} that differ from each other mainly in the way that solutions are ranked at every iteration~\cite{vazquez2012mixture}, and \textbf{MOEA/DD}~\cite{moeaddd} which is different in its decomposition technique . 
 
 \textbf{NOVA Optimisation Strategy: }as NOVA optimiser is built on a genetic (EA) strategy, it takes the following main steps in each iteration $t$, see Algorithm 1:\begin{enumerate}
		\item Each of the agents in $Ag$ and the regulative agent $r$ carry out their actions while applying the relative norms to these actions. These actions produce a new global state $s_t$. [Lines 3-6]
		\item The regulative agent $r$ performs its actions $A_r$ on $s_t$ considering the current norms $N$. [Line 7]
		\item The regulative agent $r$ uses the new global state to perform the optimisation process using a multi-objective optimiser $MOEA$ and produces the new set of norms $N$ based on the optimised set of values $V$. [Line 8]
		\item The new set of norms $N$ is communicated to all the agents in $Ag$. [Line 9]
	\end{enumerate}
  \vspace{-0.2cm}
	\begin{algorithm}
		\caption{NOVA Optimisation Strategy}
		\begin{algorithmic} [1]\label{alog1}
			\State\textbf{Input:} $N$, $V$, $s_0$, $MOEA$
			\ForEach{t} 
			\ForEach{$ag_i \in Ag$}
			\State $s_t \Leftarrow ag_i.act(N_{ag_{i}}, A_{ag_{i}},s_t)$
			\EndFor
			\State $s_t \Leftarrow r.act(N,A_{r},s_t)$
			\State $N \Leftarrow r.optimise(s_t, V, N, MOEA)$
			\State $r.inform(Ag, N)$
			\EndFor
			\State $N^{*} \Leftarrow N $
		\end{algorithmic}
	\end{algorithm}
 \vspace{-0.4cm}
	\subsection{Reasoning Process}
	\label{subsec:reasoningProcess}
The multi-objective optimisation process produces the Pareto Front set of solutions $PF_{known}$, and then, it sends each solution with its corresponding norms set to the \emph{Main Reasoner} (see Figure~\ref{fig:GeneralConceptualModel}) as $Sol$, where $sol_j=\{pf_j, N^{pf_j}_{ag_i}\}.$ 
Afterwards, a decentralised reasoning process takes place to produce one final optimum solution $sol_{best}$. 

\begin{figure}[!htbp]
		\centerline{\includegraphics[scale=0.28]{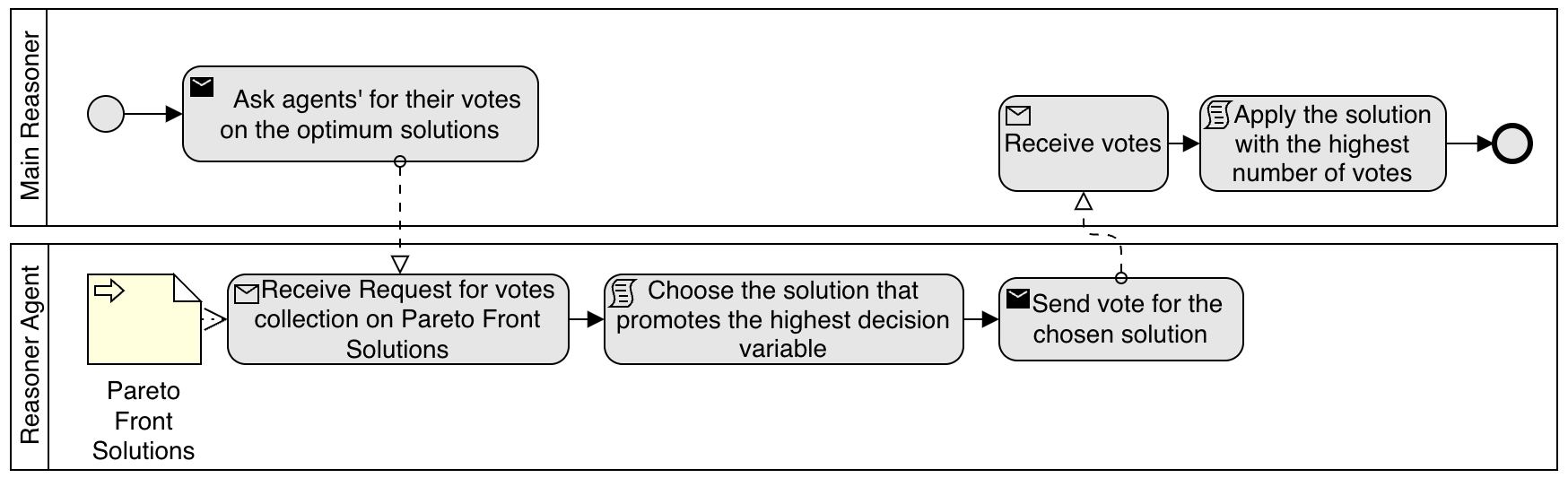}}
		\caption{Reasoning Process}
		\label{fig:ReasoningProcess}
	\end{figure}

As indicated in Algorithm~\ref{alog2} and Figure~\ref{fig:ReasoningProcess}, the reasoning process starts by running the $mainReasoner()$ after receiving the Pareto Front set and its corresponding parametric norms, and formulating $Sol$. First, in line 3, the reasoner creates an empty list to store in it the  votes that will be collected from the regular agents. Each of the reasoning (regular) agents is asked to vote in line 4 by calling the $getVote$ method. The $getVote$ method takes as parameters the Pareto Front set of optimum solutions $PF_{Known}$, and the $N^{PF_{Known}}_{ag_i}$ parametric norm values that correspond to these solutions and belong to this agent's group. In line 10, the preferred decision variable (i.e. norm to be prioritised) is stored in $prefVar$ variable. Depending on the $prefVar$, the norms set that prompts this $prefVar$ the most is stored in $n_{sol_{best}}$. Subsequently, the solution that couples this norms set is saved as the chosen solution $pf_{best}$. Then, this solution is added to the $votes[]$ at line 5. After calculating the solution with the maximum number of votes, the main reasoner states the final chosen solution in line 7.
	
  \vspace{-0.6cm}
	\begin{algorithm}
		\caption{Reasoning Algorithms}
		\label{alog2}
		\begin{algorithmic} [1]
			
			\State\textbf{Input:}, $Sol$
			\Function{$mainReasoner()$}{}$:sol_{best}$
			\State $votes[] \Leftarrow  null$
			\ForEach{$ag_i \in Ag$}
			\State $votes.add(ag_i.getVote(Sol))$
			\EndFor
                \State $sol_{best} \Leftarrow maxVotesSol(votes)$
			\EndFunction
			\Function{$getVote$}{$PF_{Known},N^{PF_{Known}}_{ag_i}$}$:pf_{best}$
			\State $prefVar \Leftarrow ag_i.getPriorityDecisionVar()$
			\State $n_{sol_{best}} \Leftarrow getMax(prefVar, N_{sol})$ 
			\State $pf_{best}\Leftarrow getObjectiveSol(n_{sol_{best}})$
			\EndFunction
		\end{algorithmic}
	\end{algorithm} 
  \vspace{-0.2cm}
\textbf{The Voting Process: }when the regular agents $Ag$ are reasoning the best solution, they calculate the fitness $fit$ of each solution by calculating equation~\ref{eq:fitnessAgentVar}. $Wg$ is the set of weights defined for each of the agent's norms, which is created randomly, and $\sum_{i}^{N}Wg_i=1$. The weights are defined based on the preferred decision variable (norms), by assigning a higher value (such as 0.8) to the preferred variable and splitting the remaining weight (0.2) among the other variables.
  \vspace{-0.3cm}
	\begin{equation}
		\label{eq:fitnessAgentVar}
		fit(PF, Wg) = select \max \forall s \in  PF\sum_{i}^{qtdVars}Wg_i * Var^s_i
	\end{equation}
   \vspace{-0.19cm}
	
 Finally, as it is expected that different agents choose different solutions
 , the most voted solution is elected and returned as the final solution.
   \vspace{0.1cm}
	\begin{equation}
		\label{eq:voting}
         \begin{aligned}
		vote(Ag)=\sum_{a \in Agents}\text{1 if }fit(PF,Ag.Wg) > fit(PF,a.Wg), \\ 
  ag \neq a
         \end{aligned}
	\end{equation}
	\begin{equation}
		\label{eq:best}
		\max \forall Ag \in Agents, vote(Ag)
	\end{equation}
\section{Tax System Scenario}\label{taxScenario}
	For further illustration of NOVA and for evaluation we will use an adapted tax system toy scenario introduced in \cite{originalProblem}. In this scenario, the regular agents set $Ag$ will represent the set of citizens and the regulating agent $r$ will represent the government. The government collects taxes from the citizens according to their wealth group. 
	There are five wealth groups, the $1^{st}$ group represents the poorest group while the $5^{th}$ group represents the richest group. A percentage of the citizens do not pay taxes and will be considered as evaders. However, if they were caught by the government they will be punished and will pay the evaded payment in addition to extra fines. In case they do not have sufficient funds only the available money is collected to avoid getting the citizen into debt. After the taxes and fines are collected a 5\% will be considered as a fixed interest rate that is added to the total collected amount. Then, the total collected money $cr$ will be redistributed back to the citizens depending on their wealth group. Initially (i.e. before simulation), The wealth of each citizen is randomly assigned after being initialized using a random uniform distribution $U$(0,100). Then, agents are allocated to their corresponding wealth group, with a constraint that the wealth groups have an equal number of citizens. The main characteristics of the system are as follows. First, each of the citizens  has four main properties in their properties set, which describes its current state. The properties are:
	  \vspace{-0.2cm}
	\begin{itemize}
		\item Wealth $(w_i)$: it has a numerical value that represents the amount of money citizen $i$ currently has.
		\item Wealth group $(g_k)$: it represents the wealth group the citizen belongs to according to its wealth $w_i$. 
		\item Evader flag $(e)$: it reflects whether this citizen is an evader and will not pay taxes or not. 
		\item Primary Wealth $(pw_i)$: it has a numerical value that represents the wealth of the citizen $i$ at the beginning of a time-step before taking any action and before its state changes.
	\end{itemize}
	  \vspace{-0.15cm}
   
	Second, each citizen has a set of values $V_{ag_{i}}$, for simplicity in this example, citizens in the same wealth group have the same fixed set of values. In other words, each wealth group has a set of values $V_{g_{i}}$, this could represent the community values. Only it is assumed that wealth group $g_2$ does not have a value to simulate citizens with no particular values, to see how they are affected by the values encouraged by others. Third, the government has its own set of values as well $V_r$ and a set of parametric norms, which has initial randomly defined values. The norms are defined in the same manner they are stated in \cite{originalProblem}:
\vspace{-0.15cm}
	\begin{itemize}
		\item \textbf{n1} defines the tax rate $collect_j$ each wealth group is expected to pay at each time-step. The parametric set of the norm is defined as $P_{n_1}$ = $\{collect_j\}_{j=1,...5}$. The tax rate values are restricted between 0 and 1.
		\item \textbf{n2} defines the fractional percentage $redistribute_j$ each wealth group will take back from the redistribution amount at the end of each time-step. The parametric set of the norm is defined as $P_{n_2}$ = $\{redistribute_j\}_{j=1,...5}$, the values are between 0 and 1 and the sum of the fractions is constrained to be equal to 1.
		\item  \textbf{n3} defines the catch rate of evaders. This single parameter is defined as $P_{n_3}$ = $\{catch\}$. Its value is constrained to be between 0 and $1/2$ to reflect the difficulty of law-enforcement tasks.
		\item \textbf{n4} defines the extra fine defined as punishment when an evader is caught. This single parameter is defined as $P_{n_4}$ = $\{fine\}$. However, the total amount to be paid by a caught evader, which is equal to the fine plus the taxes amount, can not exceed the total wealth of the evader. 
	\end{itemize}
	\vspace{-0.15cm}
 
	The main challenge of this system is represented in the government's responsibility to optimise the parameters' sets $P_{n_i}$ of the previously defined four norms belonging to $N=\{n_1,n_2,n_3,n_4\}$, while aligning them with the values of the government, as well as the regular citizen's values. The values are defined as follows.
 \vspace{-0.15cm}
	\begin{itemize}
		\item \textbf{Value 1 (Obj1)}: the value of the government is \emph{Equality}, as it aims to treat all the citizens equally without being biased to any group. \emph{Equality} is calculated using equation \ref{eq:equality} introduced in \cite{originalProblem}. $GI(s)$ represents the Gini Index of the global state $s$. The Gini Index \cite{giugni1912variabilita} is an indicator of inequality, where $w_k$ is the wealth of agent $ag_k$ and $\overline{w}$ is the average wealth of all agents at state $s$.
  \vspace{-0.15cm}
		\begin{equation}\label{eq:equality}
			Equality=1-2.GI(s),\: with\: GI(s)=\frac{\Sigma_{i,j\in Ag}\mid w_i-w_j \mid}{2.\mid Ag \mid ^2.\overline{w}}
		\end{equation}
		\item \textbf{Value 2 (Obj2)}: the value of citizens in wealth group $g_3$ is \emph{Fairness}. The main aim of this value is to have the highest number of evaders in the wealth group $g_1$ (the poorest group). To promote the estimated probability P of evaders in $g_1$ at state $s$, and to increase fairness, equation \ref{eq:fairness} is used as suggested in \cite{originalProblem}.
  \vspace{-0.15cm}
		\begin{equation}\label{eq:fairness}
			Fairness=2.P[g_i(s)=1\mid evader_i]-1
		\end{equation}
		\item \textbf{Value 3 (Obj3)}: the value of citizens in wealth group $g_5$ is to maximise their \emph{Wealth}. The main aim of this value is to have the maximum wealth portion from the total wealth. It represents the new wealth of the citizens after an iteration takes place. Equation \ref{eq:wealth} is used for calculating the new wealth.
  \vspace{-0.15cm}
		\begin{equation}\label{eq:wealth}
			Wealth=\frac{\Sigma_{i\in g_5}w_i}{\Sigma_{j\in Ag}w_j}
		\end{equation}
		\item \textbf{Value 4 (Obj4)}: the value of citizens in wealth group $g_4$ is to maximise the \emph{Gained Amount}. This value aims to have the maximum gain portion from the common amount available for redistribution $cr$. The gained value is the difference between the citizen's new wealth $w_k$ and the old wealth $pw_k$ (Check the numerator in equation \ref{eq:gained}).
  \vspace{-0.15cm}
		\begin{equation}\label{eq:gained}
			Gained \: Amount=\frac{\Sigma_{i\in g_4}w_i-pw_i}{cr}
		\end{equation}
		\item \textbf{Value 5 (Obj5)}: the value of citizens in wealth group $g_1$ is related to the \emph{Collect Portion}. This value aims to have the minimum portion from the collect rate out of 1 (a total portion of collect rates). To inverse this to a maximisation objective we have:  \ref{eq:collect}.
  \vspace{-0.15cm}
		\begin{equation}\label{eq:collect}
			Collect \: Portion=1-Collect_{g_1}
		\end{equation}
	\end{itemize}
	
	The best alignment between the synthesised set of parametric norms and the values is achieved by maximising these 5 values (objectives).
 \vspace{-0.1cm}
  \subsection{Applying the Optimisation Process of NOVA}
	\begin{figure}[!htbp]
		\centering
		\centerline{\includegraphics[scale=0.22]{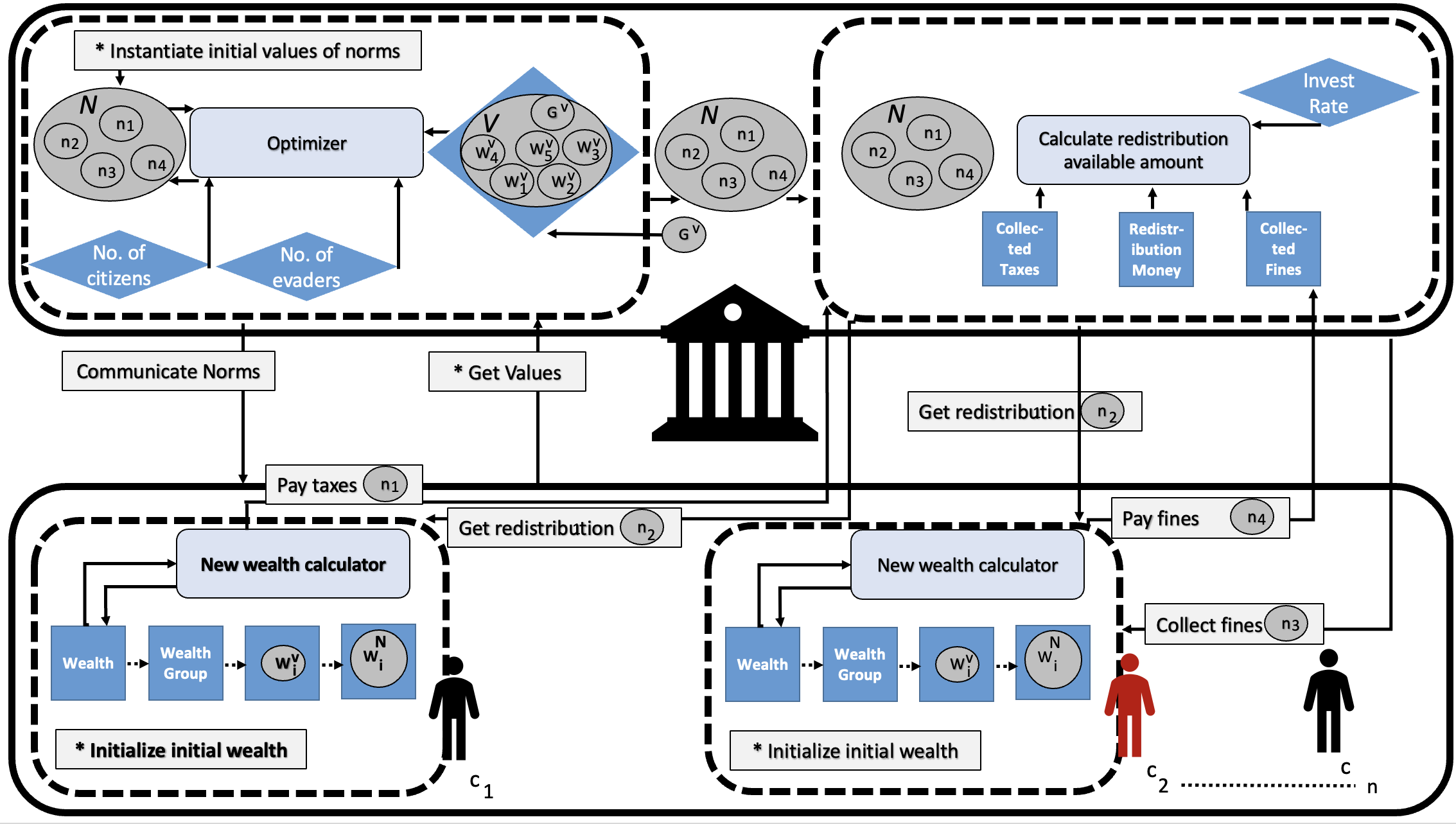}}
		\caption{Optimisation Process of NOVA in the TAX System Scenario}
		\label{fig:ConceptualModel}
	\end{figure}
	In the tax system scenario illustrated in section \ref{taxScenario}, NOVA's goal is to find the values of the parametric norms n1, n2, n3 and n4 while optimising the values in $V$: equality, fairness, wealth, gained amount and collect portion. As it is seen in NOVA's conceptual model in Fig. \ref{fig:ConceptualModel}, the system is divided into two main divisions, the divisions of the government and citizens. Evaders are represented in red as they have a different set of norms and actions than normal citizens. In this model, first, NOVA randomly initializes the norms and the wealth of the citizens, and consequently, they are assigned to their corresponding wealth groups. Second, the norms set $N$ are communicated by the government (the regulative agent) to the citizens. Third, the citizens start applying the different actions and their corresponding norms. So, normal citizens will start paying taxes according to the rate of their wealth group defined by $n1$. Then the government will start catching the evaders according to the catch value defined in $n3$. The caught evaders will pay their taxes plus the fines determined using $n4$. Afterwards, the government calculates the total amount of money available for redistribution. Subsequently, each citizen receives their portion from redistribution according to the redistribution rate defined by $n2$. Then, the citizens calculate their new wealth and move to their new wealth groups. Forth, based on $s_t$, the government uses the optimiser to decide the new values of norms by optimising the five values in $V$. This cycle is repeated until NOVA reaches a stopping condition that represents a satisfying level of the optimisation of the values in $V$.
 \vspace{-0.1cm}
 \subsection{Applying the Reasoning Process of NOVA}
 In the taxation scenario, the government will carry out the tasks of the \textit{Main Reasoner}. Accordingly, after receiving the solutions set $Sol$, it will ask the citizens to vote for the best solution $Sol_{best}$. Each citizen (i.e. each $ag \in Ag$) will randomly choose its preferred decision variable $prefVar$ to prioritise, and will choose the solution that gives the highest value for $prefVar$. If the $prefVar$ supports $n1$ or $n2$ (a parametric set of norms), the citizens will check the highest solution of the norms values that are in $P_{n1}$ or $P_{n2}$ that belongs to their wealth group. For example if an agent $ag_1$ has its $prefVar$ as $n2$ and belongs to wealth group $g_1$, it will choose from the solution that has the best value for $P_{n2}=\{redistribute_{j=1}\}$. After all the citizens vote, the solution with the maximum number of votes is set as $Sol_{best}$. 
	\section{Experimental Evaluation}
	\label{sec:experiments}
	We evaluated four algorithms NSGA-II, MOEA/DD, SPEA2, and MOMBI2 on solving both two and five objectives problems. Both problems are based on the tax scenario defined in section \ref{taxScenario}. The two objectives problem includes value 1 (Equality) and value 2 (Fairness), and the five objectives problem includes all the values. Further, we compared the results of the two-objectives scenario with a state-of-the-art work \cite{originalProblem}, as they tackle the value-alignment problem using genetic algorithm, however they handle only one value per run (i.e. do not support multi-objectives).
 
\textbf{Experimental Settings:} We used $200$ agents to represent the citizens, and a randomly chosen number of evaders in each iteration. The number of segments that represents the wealth groups was set as $5$. The investment rate was $0.05$. We used Monte Carlo Sampling during $5000$ iterations similar to \cite{originalProblem}, but in our case, Monte Carlo runs after a meta-heuristic complete execution. For this sampling the \emph{path} was defined as $10$. All meta-heuristics run for $500$ generations, with the maximum population size of $100$ for two and $210$ for five objectives. For MOEA/DD we followed \cite{moeaddd} and set $Nr=1$, $\delta=10$ and probability as $0.9$. Regarding evolutionary operators, we followed~\cite{WFGPisa}, where the SBX Crossover and Polynomial Mutation were employed and setup with distribution set as $\{n_c=20\}$ and $\{n_m=20\}$ respectively with probabilities $\{p_c=0.9\}$ and $\{p_m=1/n_p\}$, where $n_p$ is the number of decision variables in the problem.  Regarding the reasoning engine, we performed the experiment considering 200 agents.
	
	\textbf{Implementation Tools:} NOVA was coded using Java JDK 14 using jMetal 5.7~\cite{jMetal} and jMetalHyperHeuristicHelper \footnote{\href{https://github.com/vinixnan/jMetalHyperHeuristicHelper}{https://github.com/vinixnan/jMetalHyperHeuristicHelper}}.

	We discuss our results from three different perspectives. First, Hypervolume and IGD+ averages are compared to understand the performance of different algorithms in this context. Secondly, we present how is the Pareto Front for each of the meta-heuristics. Finally, we analyse the best solutions from the problem perspective.
	\subsection{Hypervolume and IGD+ comparisons}
	We employed Hypervolume \cite{jaszkiewicz2018improved}, and IGD+~\cite{IGDPlus} averages obtained from the 30 executions as the algorithms performance comparison criterion. This is necessary because MOEAs produce a set of non-dominated solutions, which makes each algorithm have a set of values to be compared. However, direct comparisons between solutions sets is difficult, therefore we need a single value that summarises the algorithm's performance. In Hypervolume, for example, we calculate the area (or volume) from each solution to a reference point. We treat them as points in the Cartesian space. For minimisation problems, we set this reference point as the worst possible. Thus, when algorithm A has a higher hypervolume value than algorithm B, it means that solutions from A are more distant to the worst point, and then the found Pareto Front provides more quality. For this purpose, first, for each problem (two and five objectives), we joined all results obtained by all algorithms, found the nadir point (worst found), necessary for Hypervolume calculation, and took the Non-dominated set in order to generate the \emph{known Pareto Front ($PF_{known}$)}, necessary for IGD+ calculation. Then, we calculated Hypervolume and IGD+ for each of the executions and generated averages for both quality indicators for each algorithm. Finally, we compared these averages using Kruskal-Wallis as the statistical test with a confidence level of $99\%$. In order to perform this, we first identified which algorithm has the best average according to the quality indicator, thus, all the other algorithms are compared to the best, generating a set of \emph{p-values}. We define an algorithm tied statistically with the best when a given \emph{p-value} is superior to the significance level of $0.01$.
	
	Table \ref{tab:2obj5obj} presents a meta-heuristic comparison for the two and five objective problems. Here the mean for 30 executions, standard deviation (\emph{std}), and  \emph{max} value among the executions are presented.
	
	\begin{table}[!htbp]
		\caption{Hypervolume and IGD+ averages for two and five objectives, highlighted values mean best results, bold values mean statistically tied results with the best. For Hypervolume, higher values are considered the best, while for IGD+ smaller is preferred.}
		\label{tab:2obj5obj}
		\centering
		\begin{tabular}{p{1.05cm}p{0.6cm}p{1.25cm}p{0.9cm}|p{1.25cm}p{0.9cm}}
			\toprule
			& & \multicolumn{2}{l|}{2 objectives} & \multicolumn{2}{l}{5 objectives}  \\
			\toprule
			& Metric &  Hypervolume &      IGD+ &  Hypervolume &      IGD+ \\
			\midrule
			MOMBI2 & mean &     0.030525 &  1.242683 &     \textbf{0.378164} &  0.086026 \\
			& std &     0.090247 &  0.354033 &     0.022794 &  0.026758 \\
			& max &     0.425940 &  1.973713 &     0.412411 &  0.137319 \\
			MOEA/DD & mean &     \textbf{0.859607 }&  \textbf{0.103377} &      \cellcolor[HTML]{DEDBDB} \textbf{0.386071} &   \cellcolor[HTML]{DEDBDB} \textbf{0.039012} \\
			& std &     0.090352 &  0.068957 &     0.009275 &  0.004931 \\
			& max &     0.974304 &  0.263218 &     0.402583 &  0.047243 \\
			NSGA-II & mean &     \cellcolor[HTML]{DEDBDB}\textbf{0.904303} &  \cellcolor[HTML]{DEDBDB}\textbf{0.056089} &     0.321709 &  0.109361 \\
			& std &     0.084679 &  0.053044 &     0.014935 &  0.014614 \\
			& max &     0.999828 &  0.239230 &     0.353739 &  0.145614 \\
			SPEA2 & mean &     \textbf{0.862924} &  \textbf{0.096932} &     0.343156 &  0.075612 \\
			& std &     0.180627 &  0.143008 &     0.009625 &  0.006417 \\
			& max &     0.999379 &  0.704847 &     0.361828 &  0.091765 \\
			\bottomrule& & 
		\end{tabular}
	\end{table}
	Regarding \emph{max} for Hypervolume for the two-objective problem, SPEA2 found the highest value. However, it also has the highest \emph{std}, which means this high value rarely occurs. NSGA-II has the best average, but when we consider both mean and std, we can see why MOEA/DD and SPEA2 results are statistically tied with NSGA-II. Regarding IGD+, also NSGA-II is the best algorithm, but this time being the best one considering \emph{std}, \emph{mean} and \emph{max}. Finally, we can clearly see that all of these algorithms except MOMBI2 can be a good option for solving this two-objective problem.
	
	For the five objective problems comparison, the scenario is completely different. MOMBI2 was a terrible algorithm for two objectives, but here it found the best \emph{max} value regarding to Hypervolume. In terms of \emph{std} and \emph{mean} MOEA/DD is the best algorithm.  These results made MOMBI2 and MOEA/DD the best algorithms considering Hypervolume. However, in terms of IGD+, MOEA/DD stands as the best algorithm with a better IGD+ average with a statistical difference. Moreover, it also had the smallest \emph{std} and \emph{max} values.

\begin{figure}[!htbp]
		\centerline{\includegraphics[scale=0.2]{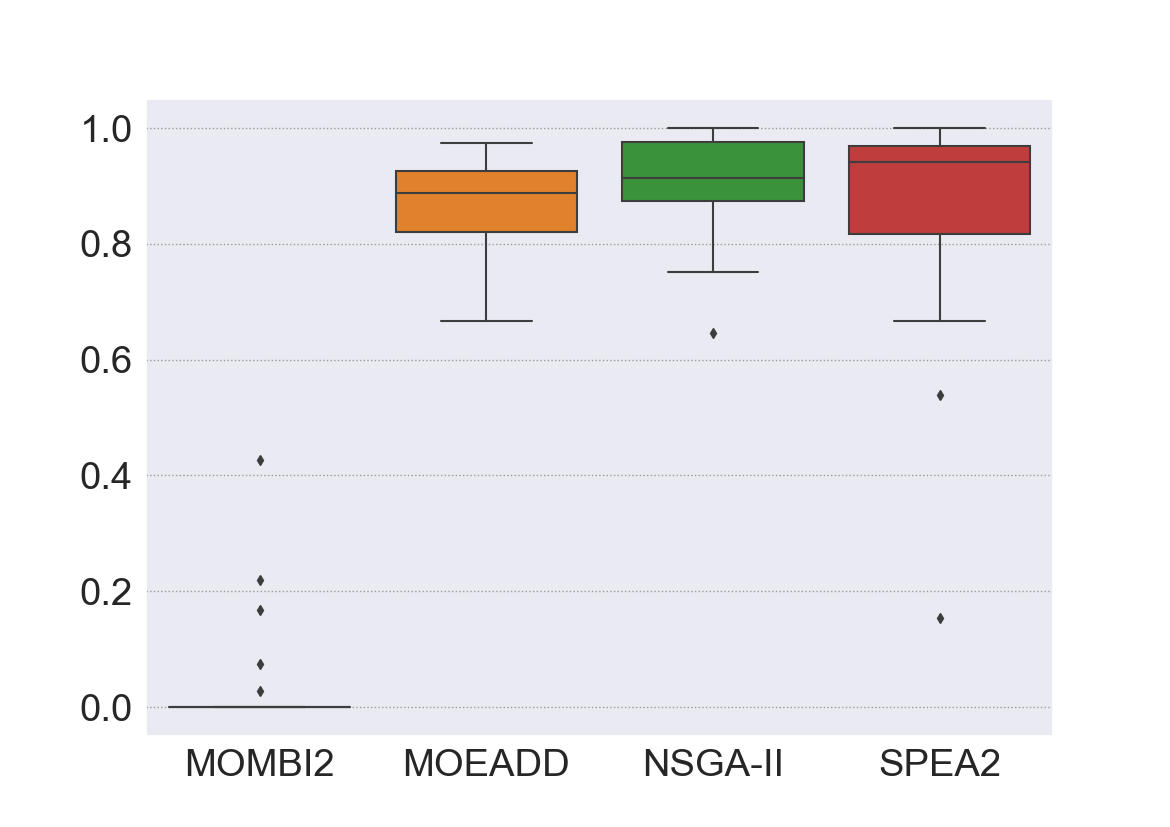}}
		\caption{Hypervolume box-plot for two objectives}
		\label{img:hypervolume2obj}
	\end{figure}

\begin{figure}[!htbp]
		\centerline{\includegraphics[scale=0.2]{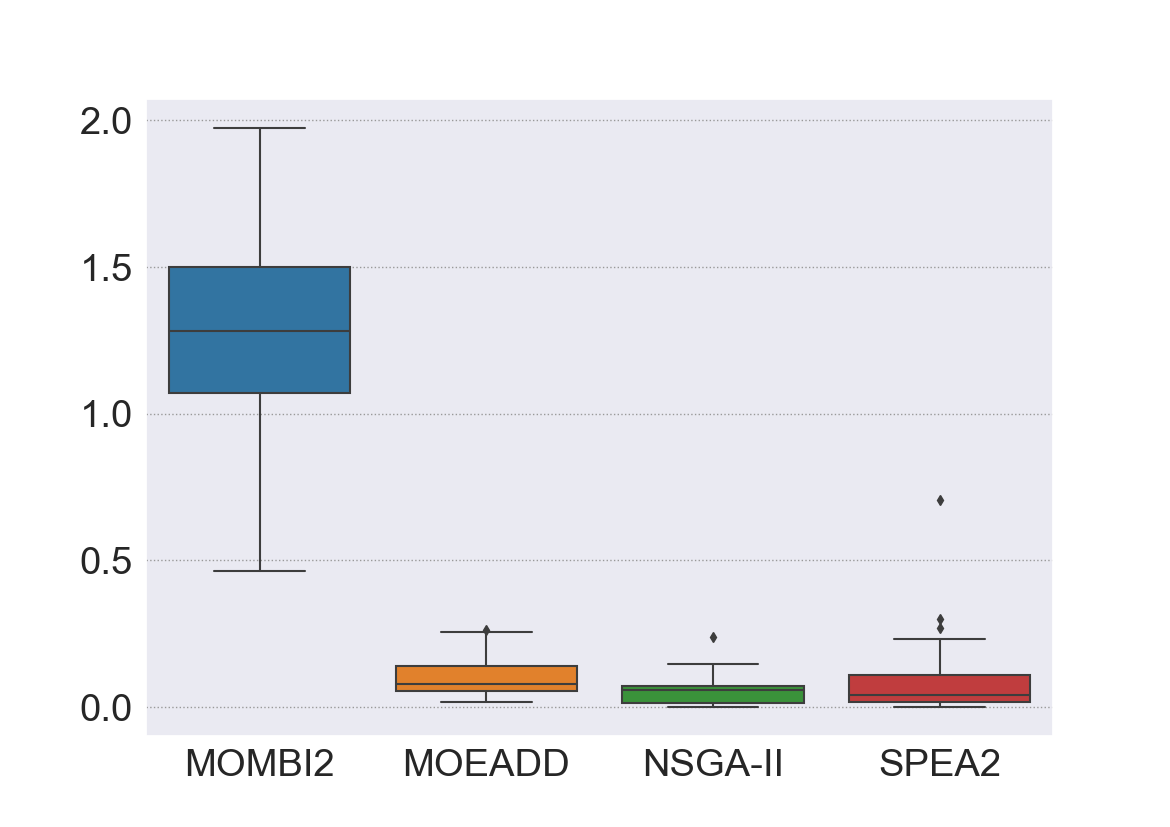}}
		\caption{IGD+ box-plot for two objectives}
		\label{img:igdp2obj}
	\end{figure}

	Figures \ref{img:hypervolume2obj} and \ref{img:igdp2obj} present, respectively, box-plots for Hypervolume and IGD+ for the two-objective problem. Basically, these figures represent visually the same results shown in Table \ref{tab:2obj5obj}. We can see how MOMBI2 is the algorithm with more variance ( in Table \ref{tab:2obj5obj} regarding \emph{std}) while NSGA-II is the one with less variance. The reason for that is while NSGA-II usually found non-dominated solutions, MOMBI2 found dominated solutions making it sometimes having near to zero Hypervolume values. This is also clear when we analyse considering IGD+ where MOMBI2 always has bigger values.
	\vspace{-0.35cm}
	\vspace{-0.83cm}

\begin{figure}[!htbp]
		\centerline{\includegraphics[scale=0.2]{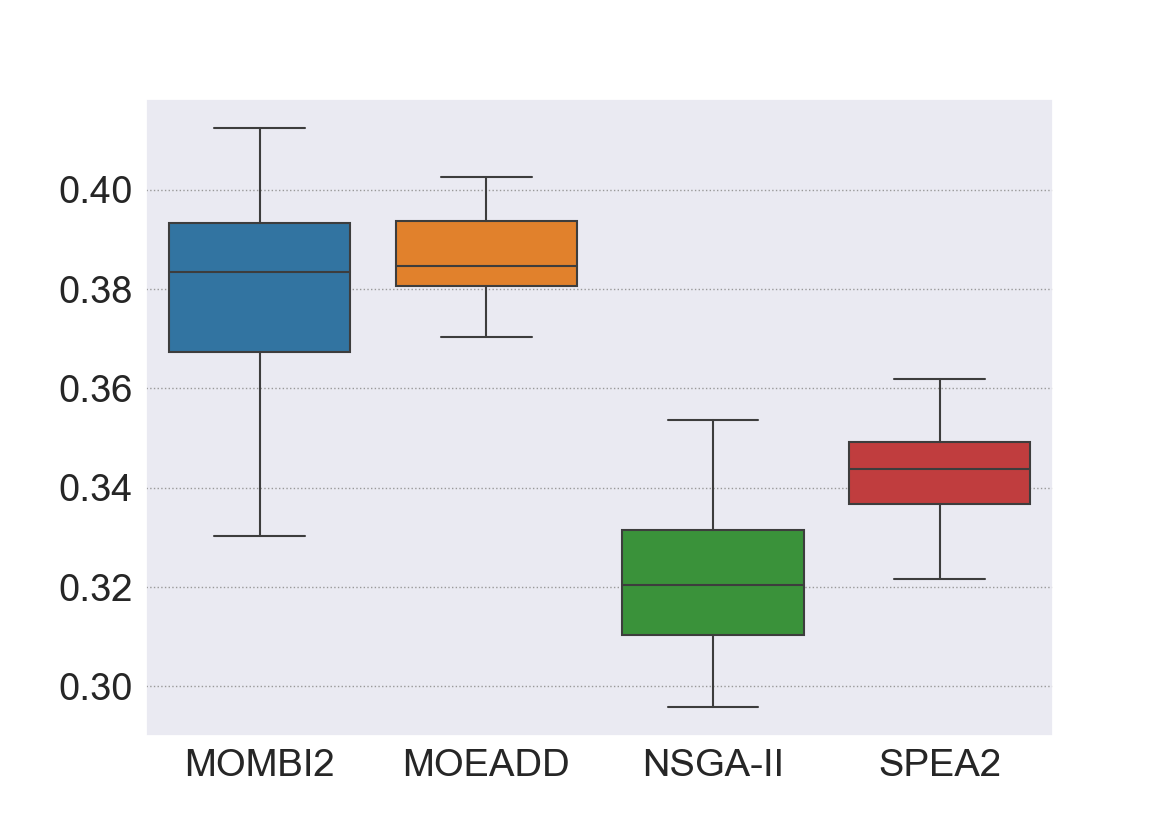}}
		\caption{Hypervolume box-plot for five objectives}
		\label{img:hypervolume5obj}
	\end{figure}

\begin{figure}[!htbp]
		\centerline{\includegraphics[scale=0.2]{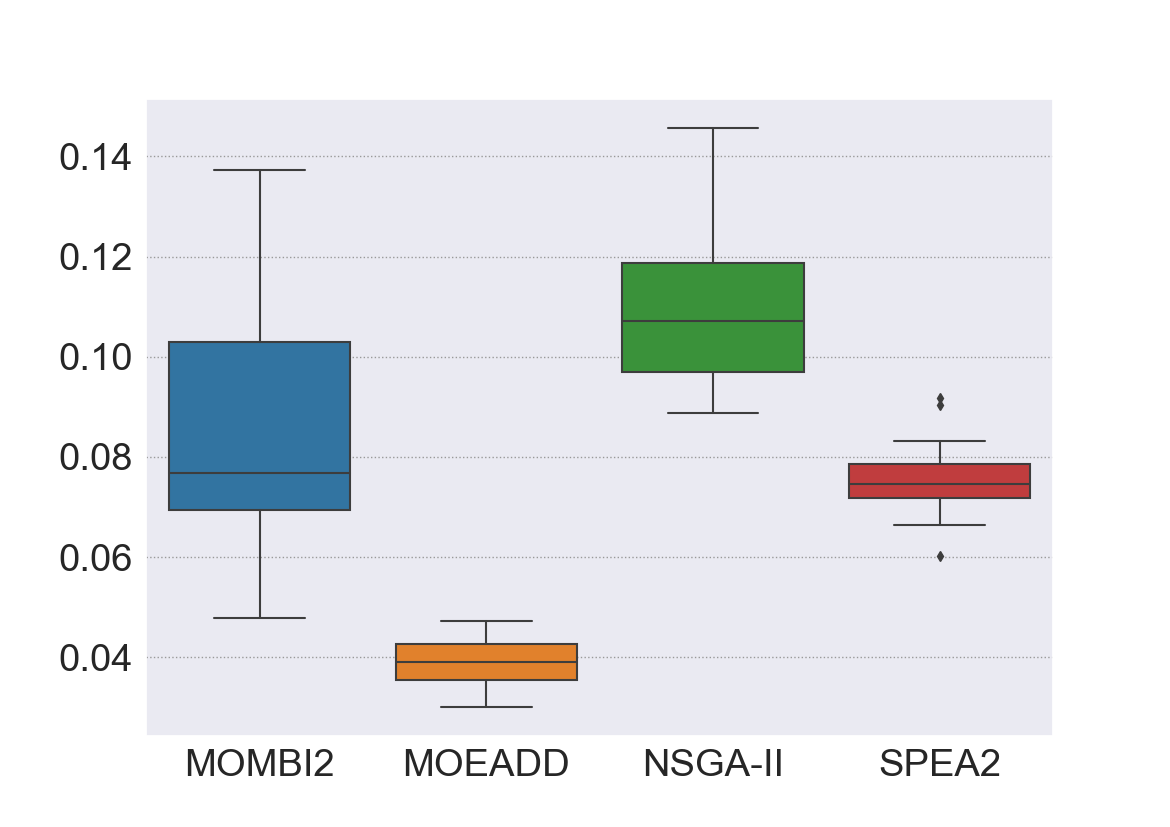}}
		\caption{IGD+ box-plot for five objectives}
		\label{img:igdp5obj}
	\end{figure}

	Figures \ref{img:hypervolume5obj} and \ref{img:igdp5obj} present respectively box-plots for Hypervolume and IGD+ considering the five-objectives problem. Basically, these figures represent visually the same results shown in Table \ref{tab:2obj5obj}. Here we can see, for Hypervolume, how MOMBI2 and MOEA/DD perform better than NSGA-II and SPEA2. However, unlike MOEA/DD, MOMBI2 has a big variance having the biggest Hypervolume values (considering one single execution) and minimal values smaller than NSGA-II and SPEA2 maximum values. MOEA/DD is more stable in terms of results, even not having the maximum value among the algorithms, which is the best choice for this problem. This is even more clear when we take into consideration IGD+ values where MOEA/DD is clearly the best-performing algorithm in all aspects.
	\vspace{-0.4cm}
	\vspace{-0.7cm}
	
		\vspace{-0.7cm}
	\subsection{Pareto Fronts Analysis}
	\label{subsec:Paretoanalysis}
	
\begin{figure}[!htbp]
		\centerline{\includegraphics[scale=0.24]{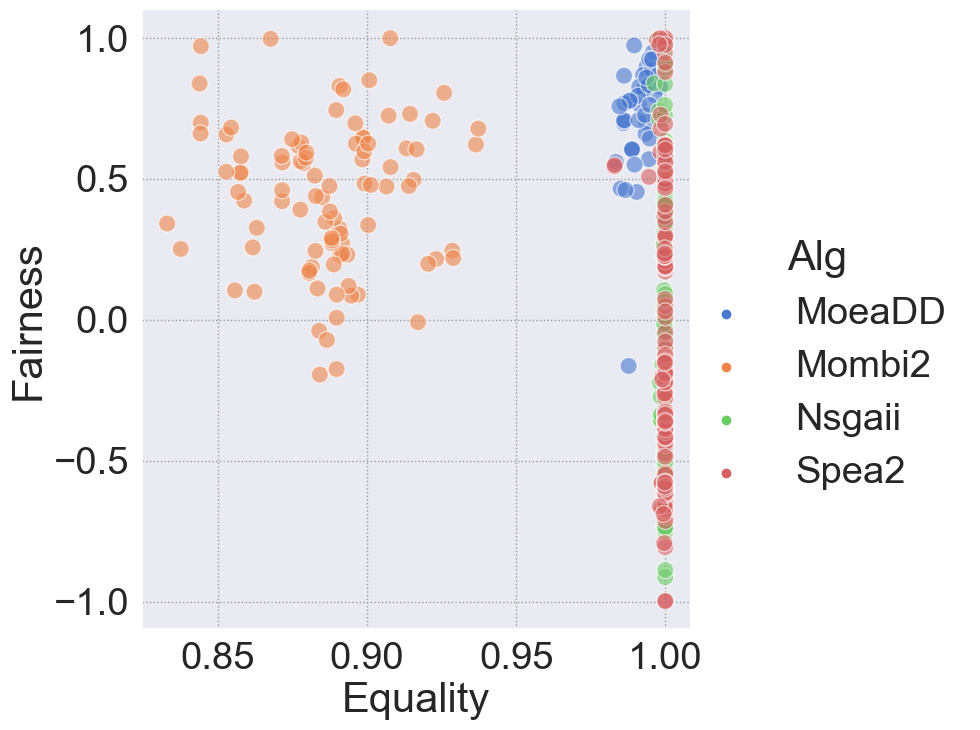}}
		\caption{Pareto Front for algorithms at one execution considering two objectives}
		\label{img:2objpf}
	\end{figure}

	In this section, Pareto Fronts generated by each algorithm are compared. Interactive graphs can be found in \href{https://htmlpreview.github.io/?https://github.com/vinixnan/PublicData/blob/master/NAO/Fronts/index.html}{\underline{\textbf{here}}}. Dominated solutions were also considered in order to provide a good view of how an algorithm can outperform others in terms of Pareto dominance. Fig.~\ref{img:2objpf} presents this for the two objective problem. Here we can see why MOMBI2 had bad Hypervolume and IGD+ results in the previous section due to the fact most of the solutions are dominated. NSGA-II also has a good amount of solutions. However, several are non-dominated, which means this algorithm has good results according to quality indicators.
	
 \begin{table*}[!htbp]
		\caption{NOVA best solution selected by the Reasoner engine for five objectives}
		\label{tab:bestsolution5obj}
		\centering
		\begin{tabular}{lccccc}
			\toprule
			Parameters                                                      & Fairness & Equality & Wealth & GainedAmount & CollectPortion \\
			\midrule
			collect = {[}70.54\%, 12.35\%, 27.00\%, 44.45\%, 37.08\%{]} &
			\multirow{4}{*}{0.8} &
			\multirow{4}{*}{0.76} &
			\multirow{4}{*}{0.24} &
			\multirow{4}{*}{1.05} &
			\multirow{4}{*}{0.29} \\
			redistribute= {[}99.79\%, 18.99\%, 89.39\%, 77.21\%, 86.50\%{]} &          &          &        &              &                \\
			catch=96.05\%                                                   &          &          &        &              &                \\
			fine=84.3634\%                                                  &          &          &        &              &               \\
			\bottomrule
		\end{tabular}
	\end{table*}
	For the five-objective problem, we split the objectives into four groups by combining objectives one and two, which are used in the two-objective problem, with one of the other objectives. Because of space limits, we decided to plot only two fronts here, and other images can be seen at \href{https://htmlpreview.github.io/?https://github.com/vinixnan/PublicData/blob/master/NAO/Fronts/index.html}{\underline{\textbf{this link}}}.
	
\begin{figure}[!htbp]
		\centerline{\includegraphics[scale=0.255]{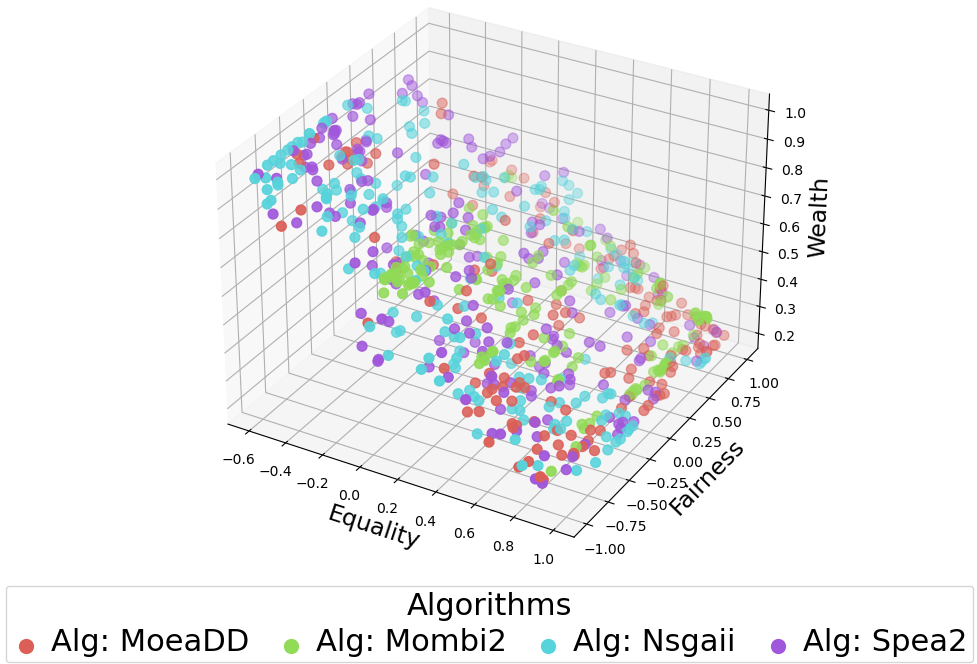}}
		
		\caption{Pareto Front for algorithms for objectives 1, 2 and 3}
		\label{img:iobj1_obj2_obj3}
	\end{figure}

\begin{figure}[!htbp]
\centerline{\includegraphics[scale=0.255]{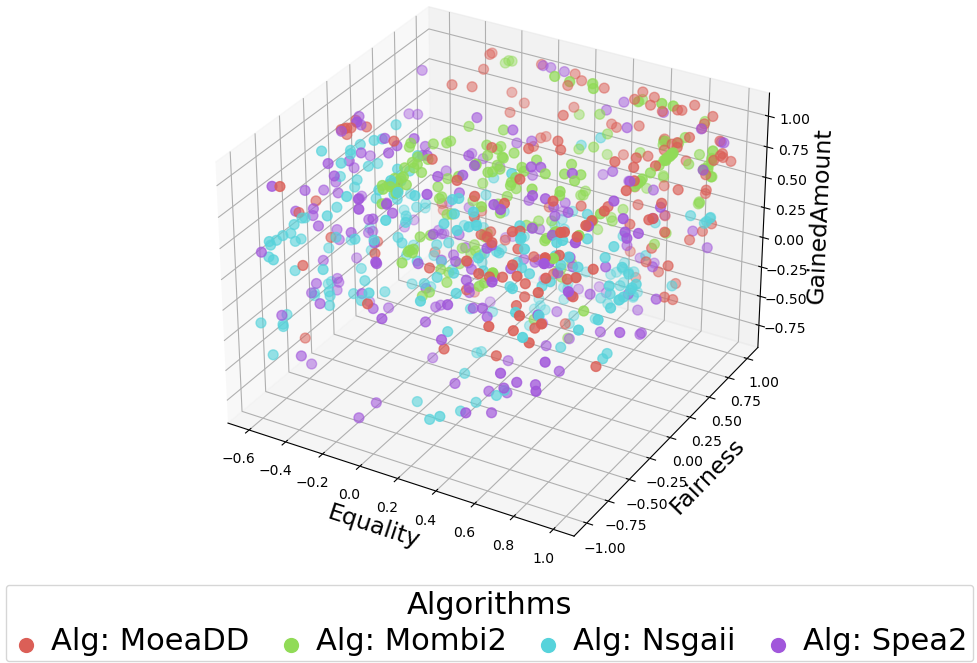}}
		
		\caption{Pareto Front for algorithms for  1, 2 and 4 objectives}
		\label{img:iobj1_obj2_obj4}
	\end{figure}

	From Figure \ref{img:iobj1_obj2_obj3}, that shows objectives 1, 2 and 3, we can view the Pareto shape, which is somehow linear and disconnected.  NSGA-II has solutions on extremes while MOEAD/D is more spread. 
 \vspace{-0.25cm}
	Figure \ref{img:iobj1_obj2_obj4} presents 3D plots for objectives 1, 2 and 4. Here we can see how MOEA/DD have solutions both spread and at extremes points. This is an excellent aspect of performance and corroborates what happened for Hypervolume values. We can also see MOMBI2 performing well here with values less spread but near to optimal. That is the reason why MOMBI2 got statistically tied results with MOEA/DD in Hypervolume, but the less diverse solutions made it have worse results when compared to MOEA/DD regarding IGD+.
  \vspace{-0.25cm}
	 \vspace{-0.1cm}
 \vspace{-0.65cm}
	\subsection{Solution analysis}
	\label{subsec:solutionanalysis}
 In this section, we analyse the results taken after the reasoning engine runs. First, in Table~\ref{tab:bestsolution2obj}, from comparing NOVA selected solution in case of the two-objectives scenario with the ones provided in ~\cite{originalProblem}, we can find that NOVA was able to achieve better results although it is aligning two values simultaneously unlike ~\cite{originalProblem}. Second, in Table~\ref{tab:bestsolution5obj}  we present the solution found by NOVA in the five-objective optimisation problem.
 \vspace{-0.1cm}
  \begin{table}[!htbp]

		\caption{Best two-objective solutions generated by NOVA according to the Reasoner compared against  solutions provided in~\cite{originalProblem} }
		\label{tab:bestsolution2obj}
		\centering
  \begin{tabular}{p{6.504cm}p{0.6cm}p{0.4cm}}
			\toprule
				Parameters& Fairness   & Equality\\
				\midrule
                    Montes and Sierra~\cite{originalProblem}& &\\
				\midrule
				 collect = {[}20\%, 29\%, 26\%, 35\%, 27\%{]}                & \multirow{4}{*}{0}    & \multirow{4}{*}{0.95} \\
			redistribute = {[}20\%, 22\%, 19\%, 26\%, 13\%{]}               &          &          \\
			catch = 44\%                                                    &          &          \\
				fine = 61\%                                                     &          &          \\
				  & & \\
			collect = {[}1\%, 30\%, 37\%, 72\%, 66\%{]}                 & \multirow{4}{*}{0.93} & \multirow{4}{*}{0.59} \\
				 redistribute = {[}2\%, 23\%, 42\%, 24\%, 9\%{]}                 &          &          \\
			 catch = 45\%                                                    &          &          \\
			 fine = 56\%                                                     &          &          \\
				\midrule
                    NOVA&&\\
				\midrule
				 collect = {[}78.45\%, 51.92\%, 60.63\%, 56.13\%, 63.15\%{]} & \multirow{4}{*}{0.80} & \multirow{4}{*}{0.86} \\
			redistribute= {[}52.75\%, 69.74\%, 47.87\%, 51.03\%, 54.30\%{]} &          &          \\
			catch= 91.72\%                                                  &          &          \\
			fine= 47.5155\%                                                 &          &          \\
				  & & \\
			collect = {[}1.96\%, 16.38\%, 35.72\%, 25.76\%, 36.19\%{]}  & \multirow{4}{*}{0}    & \multirow{4}{*}{0.93} \\
			redistribute= {[}51.82\%, 43.57\%, 52.13\%, 58.69\%, 63.23\%{]} &          &          \\
			catch= 99.88\%                                                  &          &          \\
			 fine=22.1505\%                                                  &          &         \\
				\bottomrule
			\end{tabular}
	\end{table}

 \vspace{0.1cm}
 \textbf{Discussion}\label{sec:Discussion}
After reviewing the results, it can be noted that NOVA was successfully able to:
\vspace{-0.18cm}
\begin{itemize}
    \item \textbf{Optimise multiple values regardless of their compatibility:} NOVA was able to reach this by turning the values to objectives and formalising the problem as a MOP. The produced Pareto Front set ensures having the optimum solutions for the combination of all objectives regardless of their compatibility. 
    \item \textbf{Select the best set of norms for heterogeneous groups of agents:} NOVA reached this by using the parameterised norms techniques that allow each group to have its own norms' values. 
    \item \textbf{Align independent sets of values and norms:} by defining the norms set as the decision variables set in the MOP and the values set as the objectives to be optimised, NOVA separated the concept of norms and values and had independent sets.
\end{itemize}
\vspace{-0.7cm}
	\section{Conclusion}
 \label{sec:conclusion}
	In this paper, we proposed NOVA, a multi-value promotion model that uses multi-objective evolutionary algorithms to produce optimum parametric set of norms that is aligned with multiple simultaneously shared values by heterogeneous groups of agents. 
 Moreover, it aligns values that may be incompatible such as fairness and equality, as ensuring fairness does not necessarily support equality. More importantly, we show how different algorithms can have different performance in one domain by analysing the performance of four evolutionary algorithms (NSGA-II, MOEA/DD, SPEA2, and MOMBI2). Furthermore, we enabled NOVA to produced final single optimum solution by using a decentralised reasoner.
	
 As future work, we plan to develop an online multi-value alignment model, using a hyper-heuristic approach\cite{deCarvalho2019}. Subsequently, the limitation of only including a static and limited number of pre-defined norms will be addressed by the online norm emergence techniques.
	\bibliography{ecai}
\end{document}